# R-Net: A Reliable and Resource-Efficient CNN for Colorectal Cancer Detection with XAI Integration

**Rokonozzaman Ayon**

4IR Research Cell

Department of Computer Science and Engineering

Daffodil International University, Dhaka, Bangladesh

ayon15-4393@diu.edu.bd

**Dr. Md Taimur Ahad**

School of Mathematics, Physics and Computing

Toowoomba Campus

University of Southern Queensland

MdTaimur.Ahad@unisq.edu.au

**Bo Song**

Lecturer (Industrial Automation)

School of Engineering

University of Southern Queensland

Bo.Song@usq.edu.au

**Yan Li**

Professor (Computing)

School of Mathematics, Physics and Computing

Toowoomba Campus

University of Southern Queensland

Yan.Li@usq.edu.au

**Abstract:**

State-of-the-art (SOTA) Convolutional Neural Networks (CNNs) are criticized for their extensive computational power, long training times, and large datasets. To overcome this limitation, we propose a reasonable network (R-Net), a lightweight CNN only to detect and classify colorectal cancer (CRC) using the Enteroscope Biopsy Histopathological Hematoxylin and Eosin Image Dataset (EBHI). Furthermore, six SOTA CNNs, including Multipath-based CNNs (DenseNet121, ResNet50), Depth-based CNNs (InceptionV3), width-based multi-connection CNNs (Xception), depth-wise separable convolutions (MobileNetV2), spatial exploitation-based CNNs (VGG16), Transfer learning, and two ensemble models are also tested on the same dataset. The ensemble models are a multipath-depth-width combination (DenseNet121-InceptionV3-Xception) and a multipath-depth-spatial combination (ResNet18-InceptionV3-VGG16). However, the proposed R-Net lightweight achieved 99.37% accuracy, outperforming MobileNet (95.83%) and ResNet50 (96.94%). Most importantly, to understand the decision-making of R-Net, Explainable AI such as SHAP, LIME, and Grad-CAM are integrated to visualize which parts of the EBHI image contribute to the detection and classification process of R-Net. The main novelty of this research lies in building a reliable, lightweight CNN R-Net that requires fewer computing resources yet maintains strong prediction results. SOTA CNNs, transfer learning, and ensemble models also extend our knowledge on CRC classification and detection. XAI functionality and the impact of pixel intensity on correct and incorrect classification images are also some novelties in CRC detection and classification.

**Keywords:** Colorectal cancer detection, convolutional neural network, CNN, lightweight CNN, ensemble model, SHAP, LIME, GRAD-CAM, XAI.

# 1. Introduction

The worldwide incidence of colorectal cancer (CRC) remains high because yearly diagnosis rates reach 1.8 million new cases (Cowan et al., 2022). As the second most death-causing cancer worldwide, CRC also stands among the top three cancer types (Alzahrani et al., 2021; Zhou et al., 2020). CRC caused 930,000 deaths in 2020 while generating 881,000 fatalities in 2018, according to Fadlallah et al. (2024) and deSouza et al. (2024). Researchers continue to study new treatment methods to improve CRC survival outcomes while reducing mortality rates. CRC stands as a significant worldwide public health problem because of its high death rate (Fadlallah et al., 2024). The standard diagnosis of CRC relies on histopathological examination; however, this method remains time-consuming, subjective, and requires complex analysis (Sharkas & Attallah, 2024).

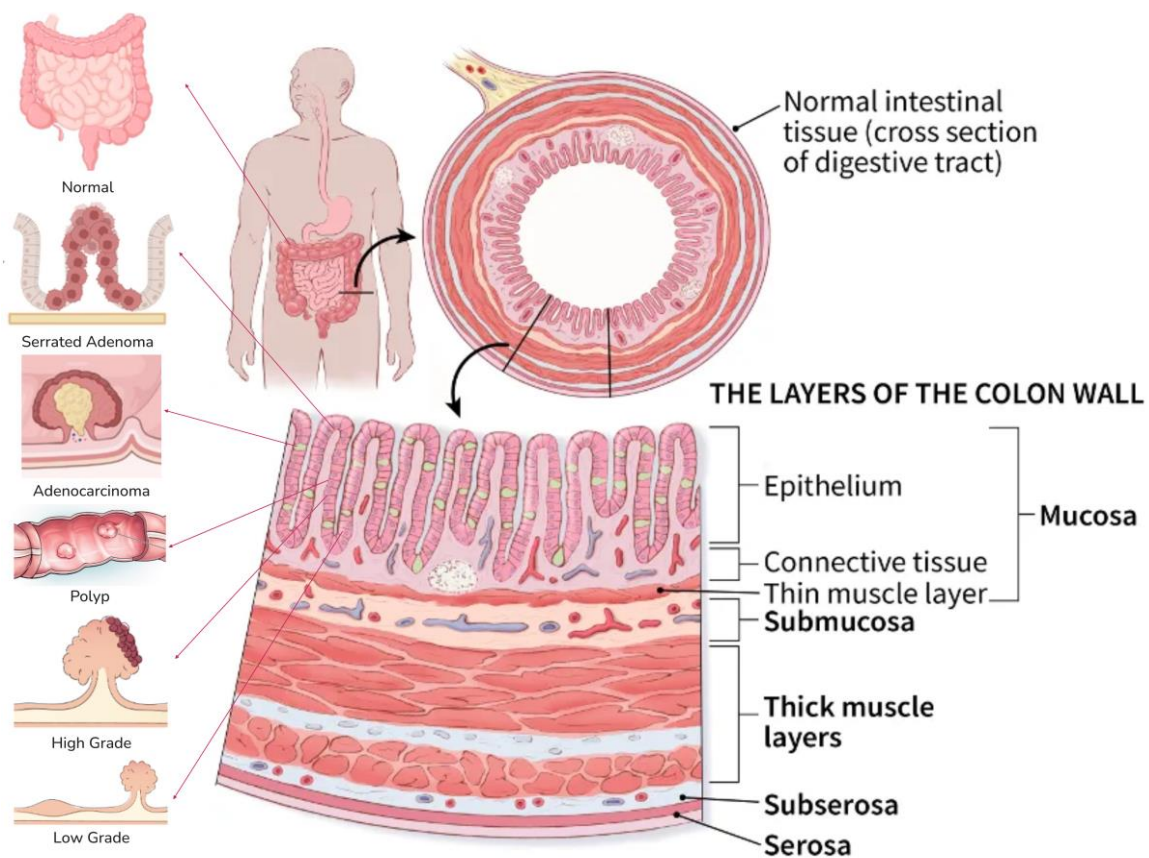

Figure 01: Visual View of CRC (amended from American Cancer Society, 2025)

In the detection and classification of CRC, Deep Learning (DL) has dramatically improved by making decisions more accurately, minimizing human error, reducing mistakes, and allowing real-time analysis in clinics (Iqbal et al., 2021). From pathology slides, the classification of cancerous tissues using DL models is very effective to achieve better accuracy in cancerous and non-cancerous tissues (Xu et al., 2020). Moreover, DL models have successfully classified polyps in colonoscopy images, and this is an essential step in CRC prevention and screening (Tanwar et al., 2022). The Deep Convolutional Neural Network (DCNN) architecture has performed very well in finding CRC-related features in histopathological images (Sarwinda et al., 2021; Shi et al., 2023). By combining the DL models with histopathological analysis, it helped to reduce the workload for diagnosis (doctors) and improve the diagnostic precision (Luo et al., 2023; Attallah et al., 2022). Among DL techniques, several approaches have been applied, including State-of-the-Art (SOTA) CNNs, modified CNNs, Transfer Learning (TL), and Ensemble models (Iqbal et al., 2022).

Despite DL's ability to detect and classify CRC, the latest DL technologies can visualize the CRC (Thakur et al., 2020). The visualizing techniques in DL systems operate under the terminology known as Explainable Artificial Intelligence (XAI). Local Interpretable Model-Agnostic Explanations (LIME), Shapley Additive Explanations (SHAP), and Gradient-weighted Class Activation Mapping (Grad-CAM) are popular XAI techniques found in the literature (Auzine et al., 2024). The combination of SHAP and LIME techniques produces both global and local explanations that work on validation and test sets for DL models (Alabi et al., 2023). Lime is basically used by a model named "black-box" to generate explanations for each prediction (Aldughayfiq et al., 2023), and SHAP is known as the most used model-agnostic method. It can be applied to any Machine Learning (ML) model to explain any model prediction. On the other hand, Grand-CAM is used to generate heatmaps for specific target classes by feature maps from the last layer of a CNN (Ghasemi et al., 2024). Mainly, it is used to show which parts of an image, or an input, are most important to predict by a model.

Several studies have been conducted on the CRC dataset for classification and identification; however, they have some limitations that can affect the reliability and applicability of these studies in a practical scenario. Numerous studies have supported the concept of lightweight CNNs and proposed novel CNN architectures that operate with fewer layers. Again, very few authors applied XAI techniques in cancer cell classifications. Some of the limitations and study gaps are mentioned below:

1. The use of imperfect ground truth data, inadequate clinical data, and insufficient

training data for validation might lead to poor performance and might question the reliability of the performance of the trained model. It could perform worse in the scenario where variability in the dataset is observed (Echle et al., 2020; Xu et al., 2020).

2. The limitation of Cancer and Non-Cancer classification in detecting CRC represents a significant disadvantage since it stands in the way of model development for alternative rare malignant conditions during colonoscopy, like sarcoma, melanoma, gastrointestinal stromal tumor, lymphoma, and carcinoid tumor (Zhou et al., 2020; Karthikeyan et al., 2024).

3. Only using a small number of images and limited patient data could create a significant issue in the trained model, and that would be called overfitting, which possibly affects the reliability of predictions and restricts model applicability (Ho et al., 2022).

4. Sarwinda et al. (2021) compared ResNet18 and ResNet50 models that might fall in comparison with other studies that have worked with many other exceptional architectures and showed their comparison analysis.

5. Prezja et al. (2024); Khazaee and Rezaee (2023) applied the ensemble strategy and proposed their ensemble model, but as the ensemble approach combines several SOTA CNNs, it has many layers. So, this issue was not addressed, and a lightweight CNN architecture was not proposed.

6. The custom model was presented by Akilandeswari et al. (2022) and Attallah et al. (2022) in their studies, but in the applicability sector, it could not reach the standard of lightweight CNN models.

7. None of the studies mentioned above have included XAI techniques like LIME, SHAP, and GRAD-CAM in their studies to explain the reasoning of their classification and misclassification.

To fill those gaps, this paper's contributions are:

1. In order to avoid overfitting issues and improve the model performance in classifying CRC, data augmentation, balancing, and some preprocessing techniques were implemented.

2. Six SOTA CNN architectures (InceptionV3, VGG16, MobileNet, ResNet50, DenseNet121, Xception) were applied on the dataset, and their performance comparison was presented.

3. Six pre-trained models with Transfer Learning were also applied to monitor the behavior of the accuracies and to show the comparison with the previous accuracies.
4. Two ensemble approaches based on three strategies, Soft-voting, Hard-voting, and Rank-based ensemble, were applied to increase the performance of the classification of CRC cells.
5. One of the main contributions was to build a lightweight Reliable Net (R-Net) model with a few layers, which achieved 99.37% accuracy with fewer resources.
6. XAI techniques like LIME and SHAP, along with Grad-CAM, were applied to make the work understandable and to make proper reasoning of classification and misclassification in CRC classification.

## 2. Literature Review

The literature covers a wide range of techniques, including colonoscopy and histological image analysis, reflecting the diversity of strategies being investigated for CRC treatment. Ahad et al., 2023; Mustofa et al., 2023; Bhowmik et al., 2024, Ahmed & Ahad, 2023; Emon & Ahad, 2024; Mustofa et al., 2024; Preanto et al., 2024; Mamun et al., 2023; Ahad et al., 2024; Mustofa et al., 2025; Preanto et al., 2024; Ahmed et al., 2023; Ahad et al., 2024; Bhowmik et al., 2023; Ahad et al., 2024; Mamun et al., 2025; Ahad et al., 2024, Ahad et al., 2024; Islam et al., 2024; Ahad et al., 2024; Ahmed et al., 2024; Ahad et al., 2024; Preanto et al., 2024; Preanto et al., 2024; Ahad et al., 2024; Ahad et al., 2024; Ahad et al., 2024; Mamun et al., 2024; Emon et al., 2023; Emon et al., 2023; Biplob et al., 2023; Ahad et al., 2023; Ahad et al., 2023; Ahad et al., 2023; Ahad et al., 2023; Ahad et al., 2023). This represents a critical advancement in cervical cancer diagnosis, enhancing the effectiveness of screening and improving early detection rates. This review highlights the transformative impact of DL on the detection and treatment of CRC by consolidating findings from several research studies.

SOTA CNNs proved the capabilities of cancerous cell detection and identification. However, it also has some limitations, for example, various layers such as concatenation, convolutional, pooling, and fully connected layers, as well as hyperparameters. Due to their large memory footprint and high computational demands (Moolchandani et al., 2021), DCNN architectures have been criticized by researchers (Thakur et al., 2023). Another researcher (Fu et al., 2024) also supported the previous researcher (Thakur et al., 2023) that CNNs have some limitations. When implementing CNNs in applied artificial intelligence, they have encountered challenges due to the complex architecture of the CNN network. To achieve a comparatively good result from DCNNs, the authors suggested lightweight CNN architectures with fewer layers, which can accurately identify the disease in cancerous images. Inspired by the success of lightweight CNNs, several studies (Thakur et al., 2023; Sun et al., 2024; Verma et al., 2024) have developed lightweight CNNs. Moreover, other methodologies are also applied in cancer cell detection, such as transfer learning and ensemble models (Xue et al., 2020).

For CRC detection, Transfer Learning (TL) is highly impactful when a large medical dataset is unavailable, as it utilizes pre-trained models for image classification. For example, to classify any cancer cell, such as colorectal polyps and cancerous tissues, TL models have been fine-tuned using CNN pre-trained models on diverse images (Alabdulqader et al., 2024; Raju et al., 2022). Techniques such as those applied in TL, partial layer freezing, and full fine-tuning help the models to focus on medical-specific features. For this reason, it continually strives to achieve better results than the pre-trained model (Davila et al., 2024; Morid et al., 2021). TL also improves the classification of benign tissues and adenocarcinomas in histopathology images (Morid et al., 2021). The Ensemble method functions as a classifier in cancer cell detection with improved accuracy than individual classification systems. It serves as an important method in many detection processes (Nanglia et al., 2022). The Ensemble model receives multiple model results from weights representing VGG19, DenseNet201, and MobileNetV2, along with other models, to enable a slow-learner algorithm for final prediction (Chugh et al., 2021). Basically, the final output is based on the cross-validated result and reduces a loss function to find optimal weights for the base model.

The remarkable performance of CRCNet has highlighted the possibility for massive DL in clinical diagnostics. This new CRC detection model was trained on a big dataset of over 464,000 pictures (Zhou et al., 2020). Using H&E-stained slides, a DL model was created to detect MSI and MMR in colorectal tumours. This model provides a faster and more affordable option to conventional molecular diagnosis (Echle et al., 2020). Effective MSI and dMMR screening for CRC was made possible by the model, which achieved an AUROC of 0.92 during development and 0.96 during validation with colour normalisation after being trained on 8,836 tumours from various nations. According to Sarwinda et al. (2021), the ResNet architecture was utilized to detect CRC in histology images and differentiate between benign and malignant instances. ResNet-50 had the best accuracy (above 80%), sensitivity (above 87%), and specificity (above 83%) across a range of test sets, demonstrating the validity of DL in the classification of CRC. To predict patient outcomes from digital tissue samples, recurrent and CNNs were combined to show that DL can extract prognostic information from tissue morphology. This approach performed better than human evaluations with an AUC of 0.69 and a hazard ratio of 2.3 (Bychkov et al., 2018). In a semi-supervised learning (SSL) technique, 13,111 whole-slide photos from 8,803 patients were utilized to train the mean teacher model (Yu et al., 2021). This approach achieved expert-level accuracy with fewer labelled patches (AUC 0.974), performing similarly to standard supervised learning in patient-level diagnosis. CRCNet, designed to enhance the identification of CRC during colonoscopy, was trained using 464,105 pictures from over 12,000 patients. It outperformed endoscopists in terms of recall rates and AUPRC values (Zhou et al., 2020). This means that CRCNet may be applied to improve CRC screening. With a high sensitivity (97.4%) and an AUC of 0.917 (Ho et al., 2022), an AI model using a Faster R-CNN architecture was created for the identification of high-risk characteristics in CRC biopsies, suggesting that it could help pathologists. An automated deep-learning approach was developed to classify colorectal polyps in histological images with 93% accuracy across five polyp types, aiding pathologists in estimating risk and enhancing screening (Korbar et al., 2022). A two-phase approach for lesion segmentation and classification was used in the development of a computer-aided diagnostic system for early CRC diagnosis utilizing CT images (Akilandeswari et al., 2022). The DCNN and residual architecture-based system showed excellent accuracy of 98.82%. In order to diagnose CRC, a two-stage classification method was suggested for separating pertinent frames from colonoscopy recordings. These frames were then classified as either neoplastic or non-neoplastic (Sharma et al., 2020). The study concluded that VGG19 was the most effective DL model for diagnosing colonoscopy images after assessing several models. To predict MSI-H in CRC using full-slide images, a DL method that integrated tumor detection and MSI classification was created (Lou et al., 2022).

## 3. Description of experimental method

This section provides the details of the hardware setup, description of the used dataset, the R-net model development, and how it will be trained for this research.

## 3.1 Hardware Specification

The experiments were conducted on a Precision 7680 Workstation equipped with a 13th-generation Intel Core i9-13950HX vPro processor and Windows 11 Pro operating system. The workstation came equipped with an NVIDIA RTX 3500 Ada Generation GPU and featured 32GB of powerful DDR5 RAM, along with a 1 TB Solid State Drive (SSD). Python V3.9 was chosen as the programming language because it worked with TensorFlow-GPU, SHAP, and LIME.

## 3.2 Dataset Description

Research data was obtained from an available public repository. Six classes composed the dataset containing Adenocarcinoma, High-Grade IN, Low-Grade IN, Normal, Polyp, and Serrated Adenoma, totaling 2228 images. A microscope instrument collected photos, which the study team stored in RGB format as PNG files. The figure displays different images that belong to each class category for this study in Figure 2.

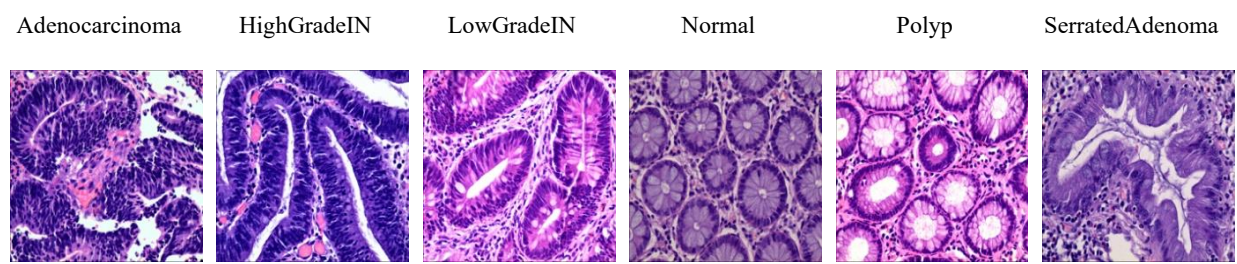

Figure 2: Samples of images used in the study.

## 3.3 Image Augmentation

In this step, the downloaded images were manually reviewed to identify class imbalances and potential issues with background color, brightness, and contrast. It was observed that the images in each class were imbalanced, a common challenge in applications such as cancer diagnosis (Johnson & Khoshgoftaar, 2019). The application of GANs helps balance the dataset by generating authentic synthetic data instances that target the minority class. A total of 4800 images were generated to balance the dataset using this technique, and the dataset distribution is shown in Figure 3.

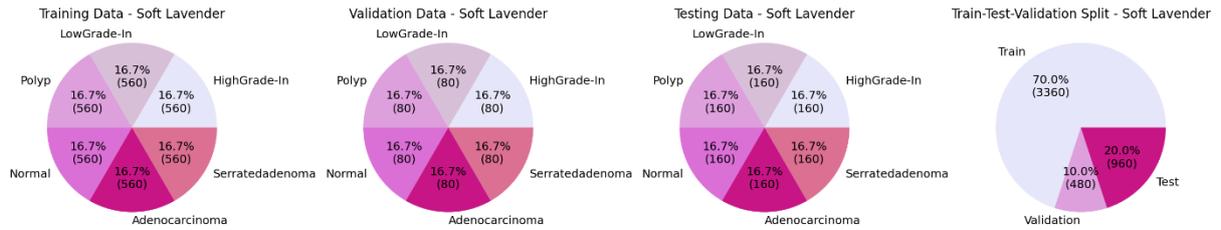

Figure 3: Distribution of images.

# 4. Results of Experiments

The researchers performed four different experiments to analyze CRC images. The analysis begins with R-Net, followed by DenseNet121, along with ResNet50, InceptionV3, Xception, MobileNetV2, and VGG16 SOTA CNNs. Then, transfer learning applied to these six SOTA CNNs. Finally, the research evaluated two ensemble models using DenseNet121 with InceptionV3-Xception and ResNet18 with InceptionV3-VGG16. The following section demonstrates experimental methodologies along with their achieved outcomes.

## 4.1 Experiment 1: R-Net development process and results

The following section explains the R-Net model together with its training process and evaluation results:

### 4.1.1 R-Net Model Development

The R-Net model was developed to find CRC cells along with their classifications within CRC image data. A set of two convolutional layers that use 64 filters begins the process before max-pooling occurs. The network adds two 128-filter convolutional layers which are followed by max-pooling before advancing to three 256-filter convolutional layers spread across more max-pooling layers. The depth of the feature map expands through successive max-pooling layers following three 512-filter convolutional layers that automatically reduce spatial dimensions. Feature extraction ends with flattening the output before passing it to two dense layers that have a fully connected structure.

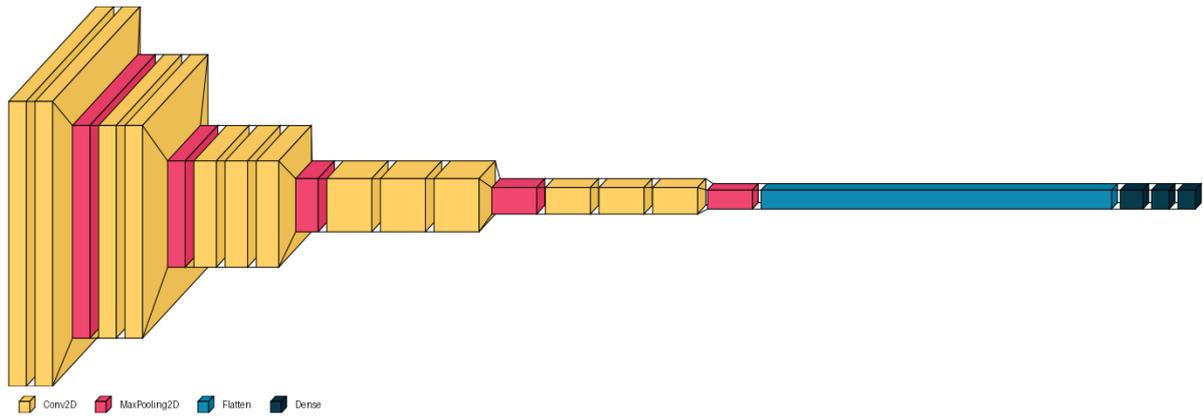

Figure 4: R-Net model visualisation

The initial dense layer contains 256 neurons, and the second dense layer has 64 neurons. Six neurons form the last layer structure because it contains every possible target class for classification purposes. The model contains 15,911,430 trainable parameters for extracting image features, enabling its use in multiclass image classification.

### 4.1.2 Training setup

The R-Net model was trained using 5-fold cross-validation. The training process extended for over 45 epochs using batches of size 32 in each fold. The *Adam optimization* algorithm was used for model optimization. The weights of the model became adjustable through gradients calculated by the algorithm, which resulted in enhanced classification performance and accuracy of CRC modalities—the selected loss function employed sparse categorical cross-entropy for data calculation. The training parameters can be found in Table 1.

Table 1: Hyperparameters of training

| Parameter | Value |
| --- | --- |
| Epochs | 50 |
| Batch size | 16 |
| Image size | (64, 64, 3) |
| Learning rate | 1.00e-04 |
| K_folds | 5 |
| Optimizer | Adam(learning_rate=LEARNING_RATE) |
| Loss Function | SparseCategoricalCrossentropy(from_logits=True) |
| Early Stopping | EarlyStopping(monitor='val_accuracy', patience=10, verbose=1, restore_best_weights=True) |

| | | |
|---|---|---|
| Learning Rate Scheduler | LearningRateScheduler( lambda epoch: LEARNING_RATE * 0.1 ** (epoch // 10) | |
| Callbacks | [early_stopping, lr_scheduler] | |

### 4.1.3 Results of the R-Net

Table 2 presents the evaluation of the R-Net model performance for each fold, which includes precision, recall, F1-score, and support. All five evaluations produced high-accuracy results through the model while maintaining low mistake rates. The model in Fold 1 achieved near-perfect precision but misclassified some instances. However, the classification performance in Fold 2 proved exceptional because the model achieved outstanding results without any significant misclassifications. Folds 3, 4, and 5 displayed outstanding performance, as misclassification was minimal. The model demonstrates exceptional capabilities in classifying different categories with high precision, thanks to its outstanding error reduction capabilities.

Table 2: Fold-Wise Classification report with epochs of R-Net

| Fold | Class | Precision | Recall | F1-Score | Support |
|---|---|---|---|---|---|
| 1 | Adenocarcinoma | 1 | 0.99 | 0.99 | 138 |
| | HighGradeIN | 0.99 | 1 | 1 | 120 |
| | LowGradeIN | 0.99 | 0.99 | 0.99 | 133 |
| | Normal | 1 | 1 | 1 | 119 |
| | Polyp | 0.99 | 1 | 1 | 125 |
| | SerratedAdenoma | 1 | 0.99 | 1 | 133 |
| 2 | Adenocarcinoma | 0.98 | 0.99 | 0.99 | 132 |
| | HighGradeIN | 0.99 | 0.99 | 0.99 | 137 |
| | LowGradeIN | 0.98 | 0.97 | 0.98 | 128 |
| | Normal | 1 | 0.99 | 1 | 131 |
| | Polyp | 0.98 | 0.98 | 0.98 | 119 |
| | SerratedAdenoma | 0.99 | 1 | 1 | 121 |
| 3 | Adenocarcinoma | 1 | 0.97 | 0.98 | 128 |
| | HighGradeIN | 0.98 | 1 | 0.99 | 137 |
| | LowGradeIN | 0.99 | 0.98 | 0.99 | 126 |

|   | Normal | 0.99 | 1 | 1 | 131 |
|---|---|---|---|---|---|
|   | Polyp | 0.98 | 0.99 | 0.98 | 124 |
|   | SerratedAdenoma | 1 | 0.99 | 1 | 122 |
| 4 | Adenocarcinoma | 1 | 0.99 | 1 | 130 |
|   | HighGradeIN | 0.98 | 1 | 0.99 | 124 |
|   | LowGradeIN | 1 | 0.98 | 0.99 | 123 |
|   | Normal | 1 | 1 | 1 | 126 |
|   | Polyp | 0.98 | 0.99 | 0.98 | 122 |
|   | SerratedAdenoma | 1 | 1 | 1 | 143 |
| 5 | Adenocarcinoma | 0.98 | 0.98 | 0.98 | 112 |
|   | HighGradeIN | 0.98 | 1 | 0.99 | 122 |
|   | LowGradeIN | 0.98 | 0.97 | 0.97 | 130 |
|   | Normal | 1 | 1 | 1 | 133 |
|   | Polyp | 0.99 | 0.98 | 0.98 | 150 |
|   | SerratedAdenoma | 1 | 1 | 1 | 121 |

The model's precision level becomes noticeable through visualization in the confusion matrix presented in Figure 5. In Fold 1, the model performed well with very minimal misclassification errors, and Fold 2 achieved better accuracy by successfully separating challenging class samples. The model reached exceptional levels of classification in Folds 3 through 5 because errors reached virtually zero during these runs. The model successfully differentiates multiple categories, exhibiting high precision and recall, which proves its effectiveness in minimizing misclassification errors and ensuring reliability.

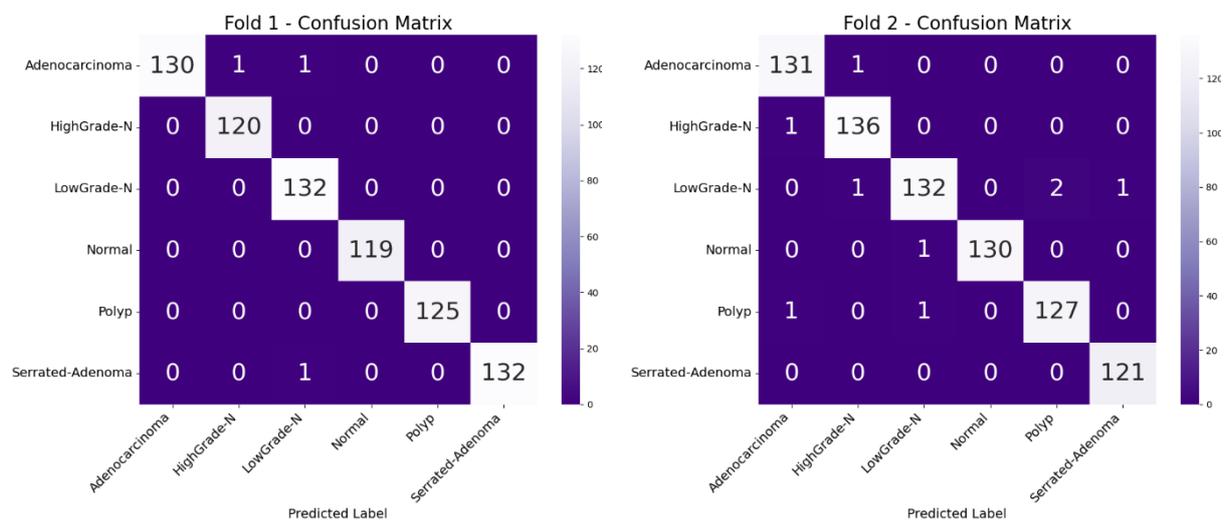

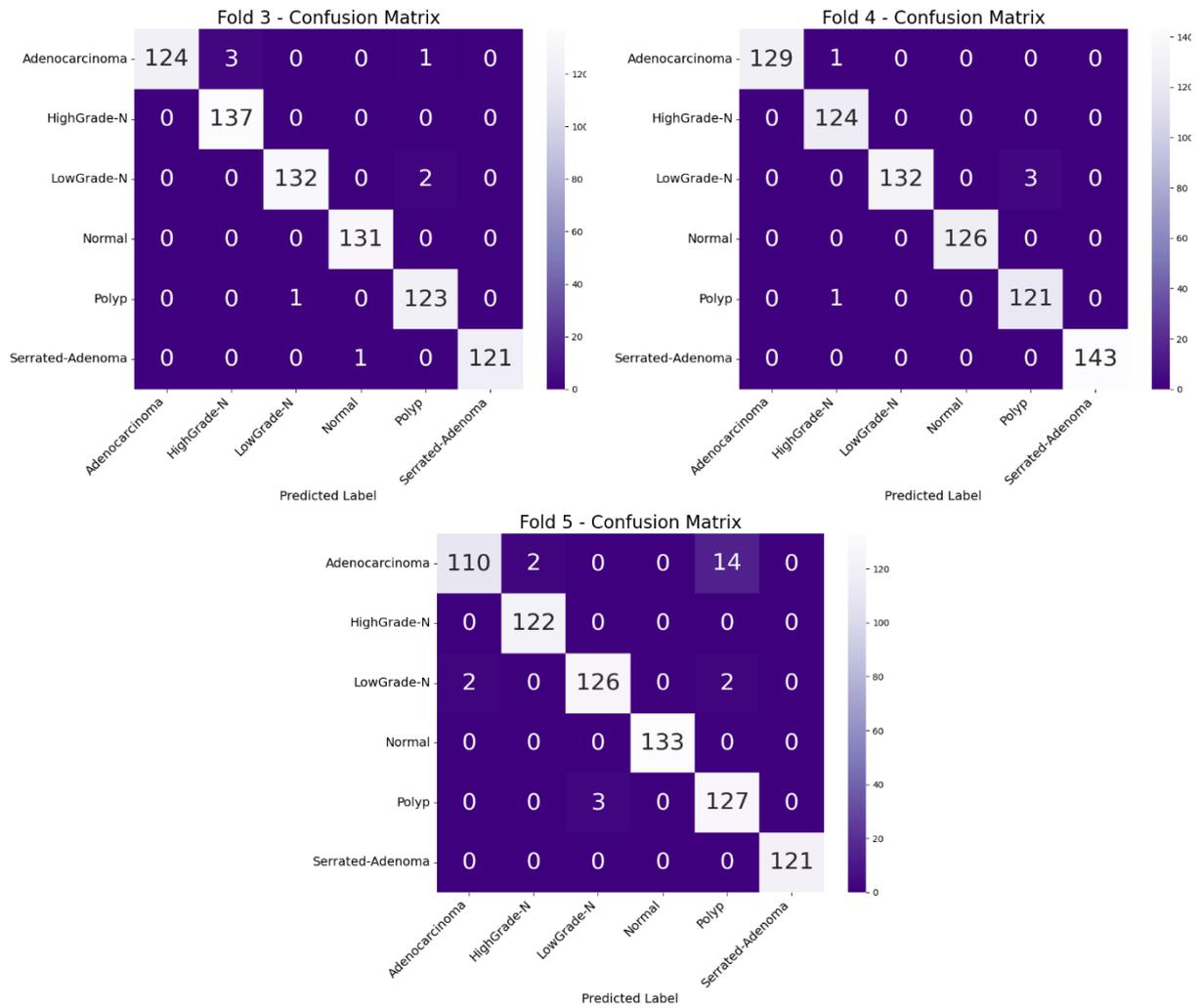

Figure 5: Fold-wise confusion matrix of R-Net.

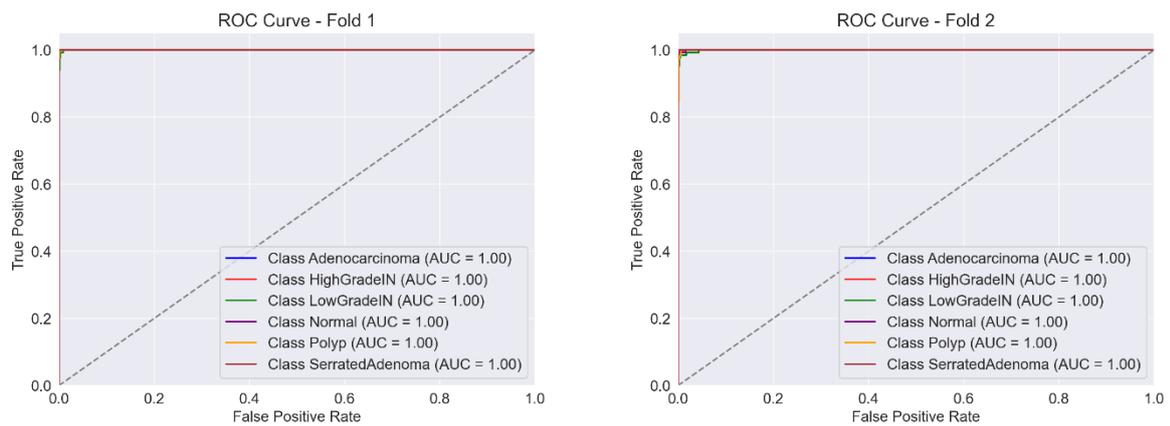

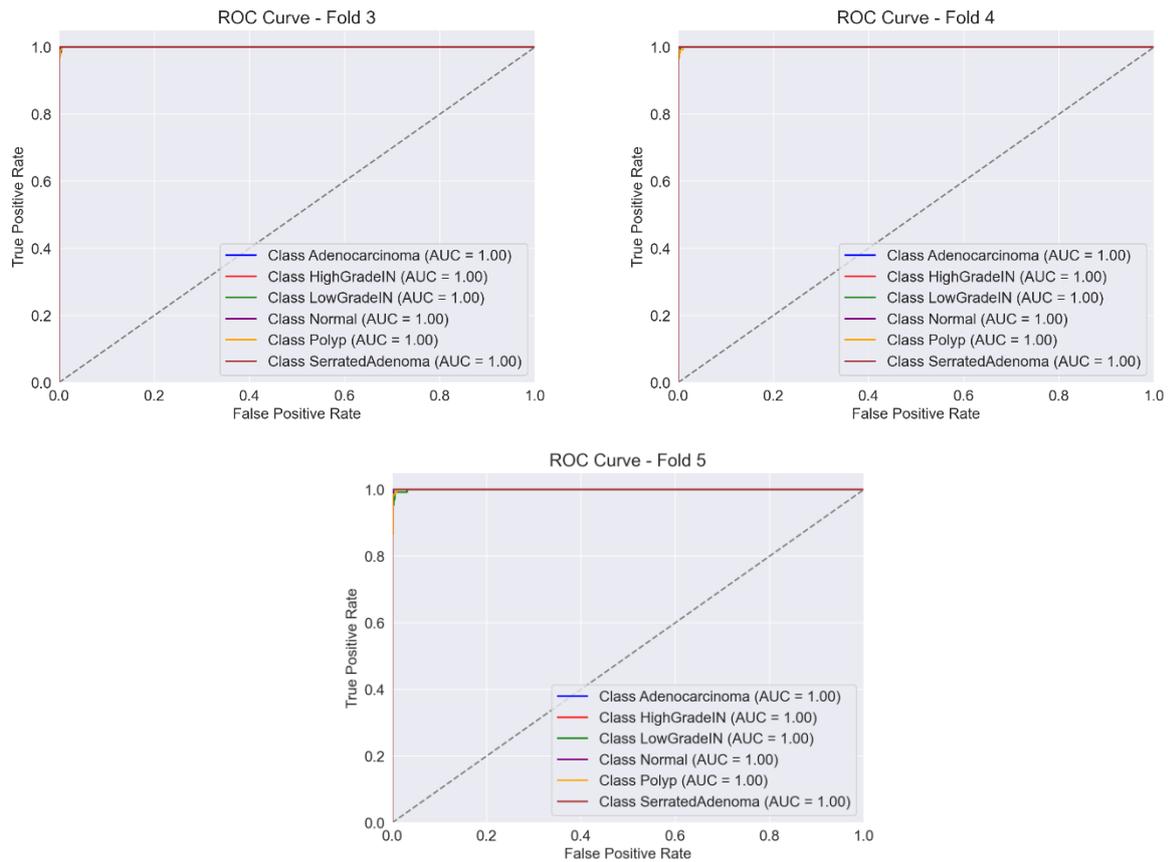

Figure 6: Fold-Wise ROC-Curve of R-Net.

A comparison of different ROC curves appears in Figure 6 based on five-fold cross-validation techniques. The evaluation methods of Fold 1 show both high accuracy in correctly identifying cases and correctly misclassified cases while minimizing false positive errors. The updated Fold 2 enhances the model with a denser curve design. The performance accuracy of the model becomes evident through near-perfect ROC curves that appear in Folds 3 through 5. The reliability and robustness of the R-Net model are evident in these achieved results in multi-class classification.

Figure 7 displays training and validation accuracy and training and validation loss data for the five R-Net model folds. The plot illustrates both training accuracy and validation accuracy rates, alongside a decreasing training loss and sustained low validation loss, which signifies outstanding model performance and avoids overfitting occurrences. The model demonstrates reliable performance and strong generalization capabilities across all folds, as indicated by these results.

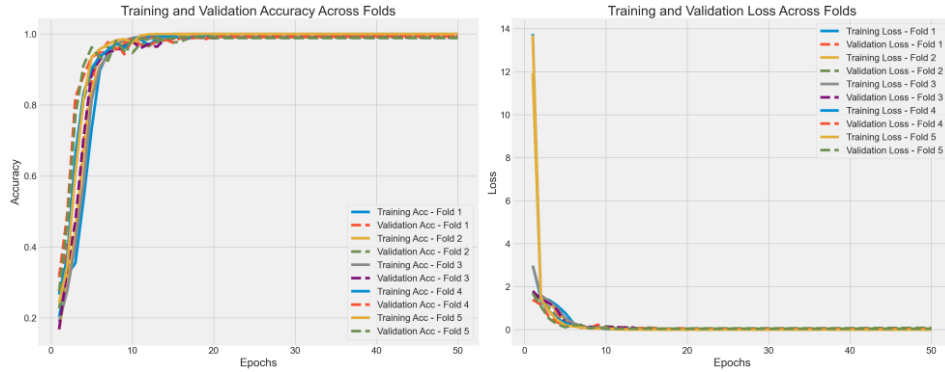

Figure 7: Training and Validation across all folds.

Additional performance evaluation of the R-Net model generated a confusion matrix based on the test dataset. Figure 8 presents the model classification results, which show accurate predictions among different categories. The model demonstrated robustness and reliability through the match between the classification matrix and its high-accuracy assessment. A small sample misidentification demonstrates the model's efficient generalization effectiveness, which qualifies it for practical utilization.

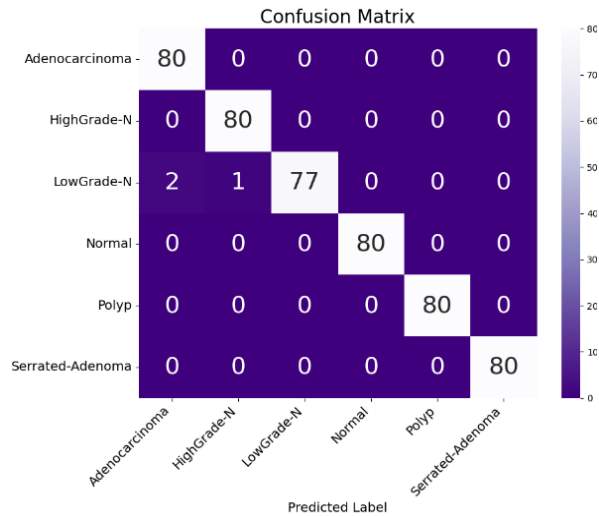

Figure 8: Confusion Matrix of the R-Net Model on the Test Dataset

The performance metrics for the R-Net model appear in Table 3 for training, validation, and the test datasets.

Table 3: Model Performance Metrics on Training, Validation, and Test Sets

| Dataset | Loss | Accuracy |
| --- | --- | --- |
| Train | 1.06e-07 | 100% |
| Validation | 0.012 | 99.79% |
| Test | 0.0275 | 99.37% |

The model learned the training data efficiently, achieving a minimal training loss value of 1.06e-7 (0.00000010617) along with perfect accuracy of 100%. The validation loss shows minimal value (0.0120) alongside a high accuracy of 99.79% which indicates strong generalization to new data points. The test data shows both low loss at 0.0275 and accuracy at 99.37% which strengthens the reliability and robustness of the model. The model demonstrates excellent potential for practical application, as it achieves high classification accuracy while minimizing errors.

R-Net delivers outstanding performance in its combined classification metrics by achieving a 99% accuracy across every category. The model achieves equal and highly effective results across precision, recall, and F1-scores, with values of approximately 0.99. The model achieves strong performance based on specific classification results, which show that the Normal, Serrated Adenoma, and Polyp categories achieve scores close to 1.00. The evaluation of Adenocarcinoma, High-Grade IN, and Low-Grade IN cancerous cell types through R-Net shows that the model achieves precision, recall, and F1 scores between 0.97 and 0.99. The model demonstrates reliability through its consistent performance, as shown by macro and weighted averages across the entire dataset.

The confusion matrix exhibits the R-Net's high accuracy. Among 4800 images, R-Net correctly detected and classified 4797 images. However, only 3 LowGradeIN instances were misclassified as 2 Adenocarcinoma and 1 HighGradeIN, while Normal and Polyp showed no misclassification errors. The confusion matrix confirms that the model successfully reduces false positive results.

The fold-wise accuracy and loss curves deliver details about how well the model performs throughout training with validation intervals. The model's learning process exhibits significant improvement in accuracy, ultimately achieving high levels of model ability in terms of generalization and learning. Successful model operation maintains stable learning efficiency, but error reduction appears firmer because of decreasing loss curve values. The R-Net model achieved high accuracy along with minimal loss values during training sessions.

## 4.2    Experiment 2: SOTA CNN performance on CRC

An examination of six SOTA CNNs is conducted according to the taxonomy system of Khan et al. (2020). The models organized into five categories include Depth-based CNNs (InceptionV3), Multi-Path-based CNNs (ResNet50, DenseNet121), and Width-based Multi-

Connection CNNs (Xception), Depthwise Separable Convolutions (MobileNet), along with Spatial Exploitation-based CNNs (VGG16). The selection of these models was done to provide deep insight into which CNN produces the best results for CRC image classification. The performance of CNNs during this task was evaluated using three different optimizers: Adam, Adamax, and RMSprop.

Table 4: Performance comparison of SOTA CNNs and optimizers

| Model | Epoch (Adam) | Accuracy (Adam) | Epoch (Adamax) | Accuracy (Adamax) | Epoch (RMSprop) | Accuracy (RMSprop) |
|---|---|---|---|---|---|---|
| DenseNet121 | 29 | 0.9993 | 29 | 0.9965 | 23 | 1 |
| ResNet50 | 28 | 0.9694 | 27 | 0.9722 | 29 | 0.9917 |
| InceptionV3 | 22 | 0.9944 | 37 | 0.9625 | 24 | 0.9993 |
| Xception | 14 | 1 | 14 | 1 | 14 | 1 |
| MobileNet | 50 | 0.9583 | 46 | 0.8465 | 32 | 0.9257 |
| VGG16 | 11 | 0.1667 | 30 | 0.9674 | 24 | 0.9944 |

The performance metrics of six state-of-the-art CNNs are presented in Table 4 under Adam, Adamax, and RMSprop optimizer conditions. The highest accuracy of 99% is achieved by DenseNet121 using Adam (0.9993) and RMSprop (1.0000), which required 29 epochs alongside 23 epochs. The performance of ResNet50 demonstrates reduced accuracy when Adam reaches 0.9694, while Adamax reaches 0.9722, and RMSprop generates 0.9917 accuracy, yet optimizers fail to affect accuracy measurements noticeably. The combination of RMSprop (0.9993) and Adam (0.9944) delivers superior results than Adamax (0.9625) for InceptionV3. Within 14 epochs, the Xception model achieves a perfect accuracy score of 1.0000 when combined with any of the optimizers. The accuracy of MobileNet is lower when using Adamax (0.8465), while both Adam (0.9583) and RMSprop (0.9257) outperform it in terms of accuracy. Adam produces subpar results for the VGG16 model, with an accuracy rating of only 0.1667, whereas Adamax achieves 0.9674 accuracy and RMSprop delivers 0.9944 accuracy. The results demonstrate that Adam and RMSprop provide stable performance metrics, although Adamax shows reduced operational efficiency, mainly when used in MobileNet and InceptionV3 models. The accuracy-epoch relationship between optimizers and models appears in Figure 9 through a scatter plot representation.

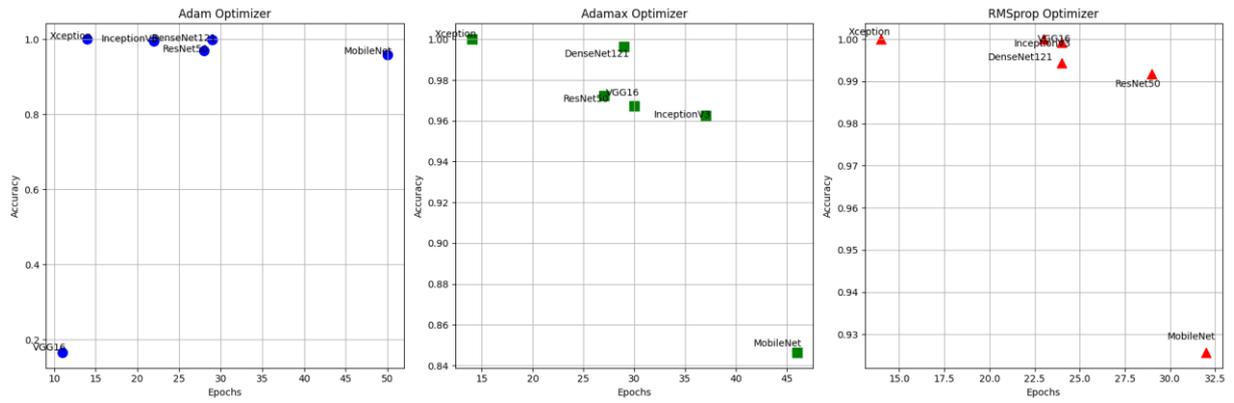

Figure 9: The training and validation accuracy of the original CNNs.

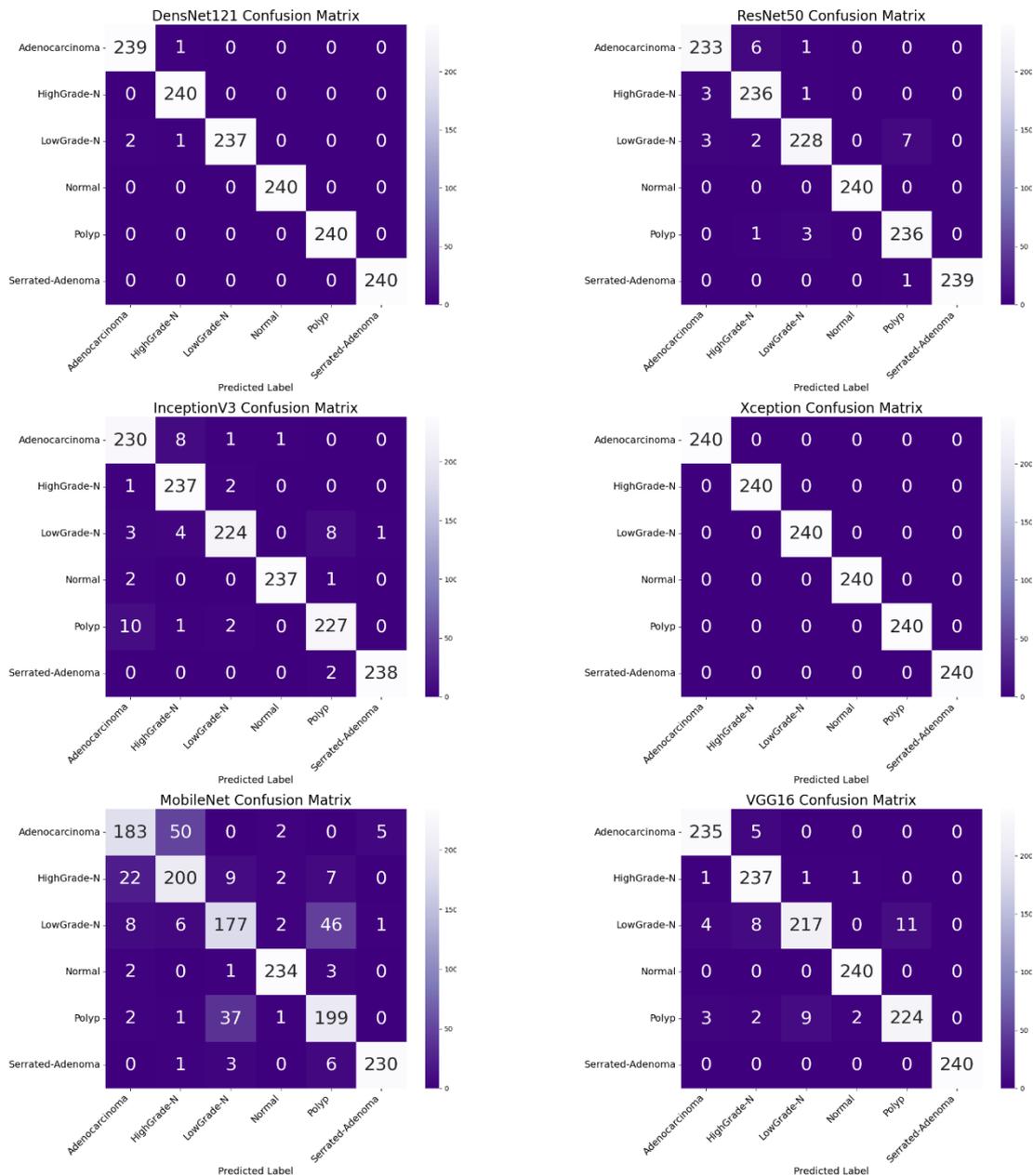

Figure 10: Confusion matrices of Transfer learning

The confusion matrix in Figure 10 shows that Xception and DenseNet121 have lower Type 1 and Type 2 errors, which indicates better classification performance. The performance of InceptionV3 falls within the middle range, as it generates both acceptable false and true negatives and positives. The misclassification rates of VGG16 remain high, especially in LowGrade-IN and Polyp, which makes this model less dependable than the others. The misclassification rates for ResNet50 are significantly elevated in classifications of Adenocarcinoma and Low-Grade IN, resulting in numerous incorrect positive and negative predictions. MobileNet demonstrates the worst capability among these models based on its classification errors in HighGrade-IN, LowGrade-IN, and Polyp.

## 4.3 Experiment 3: SOTA CNNs Transfer Learning Performance on CRC

An evaluation of six SOTA CNNs transfer learning architectures occurs during this experiment. The experiment relies on the same classification system from Experiment 1, using SOTA CNN for equal model comparison throughout. Evaluation takes place through an analysis of training, validation, and testing accuracy from distinct optimization methods.

Table 5: Performance comparison of SOTA CNNs; Transfer learning and optimizers

| Model | Epoch (Adam) | Accuracy (Adam) | Epoch (Adamax) | Accuracy (Adamax) | Epoch (RMSprop) | Accuracy (RMSprop) |
|---|---|---|---|---|---|---|
| DenseNet121 | 5 | 0.7962 | 5 | 0.7456 | 5 | 0.7883 |
| ResNet50 | 5 | 0.5677 | 5 | 0.4385 | 5 | 0.4985 |
| InceptionV3 | 5 | 0.8835 | 5 | 0.7244 | 5 | 0.861 |
| Xception | 5 | 0.63 | 5 | 0.4125 | 5 | 0.5902 |
| MobileNet | 5 | 0.8769 | 5 | 0.704 | 5 | 0.8077 |
| VGG16 | 5 | 0.7942 | 5 | 0.6342 | 5 | 0.7025 |

The accuracy data demonstrate that DenseNet121 achieves high accuracy rates using multiple optimizers, with Adam reaching 79.62%, Adamax reaching 74.56%, and RMSProp reaching 78.83% (Table 5 and Figure 11). The results show that MobileNet maintained consistent performance, achieving 87.69% accuracy with Adam, 70.40% accuracy with Adamax, and 80.77% accuracy with RMSProp. The accuracy results show that ResNet50 achieves the worst performance in transfer learning, with 56.77% accuracy from Adam, 43.85% from Adamax, and 49.85% from RMSProp, even though all optimizers required five epochs, which indicates optimization difficulties.

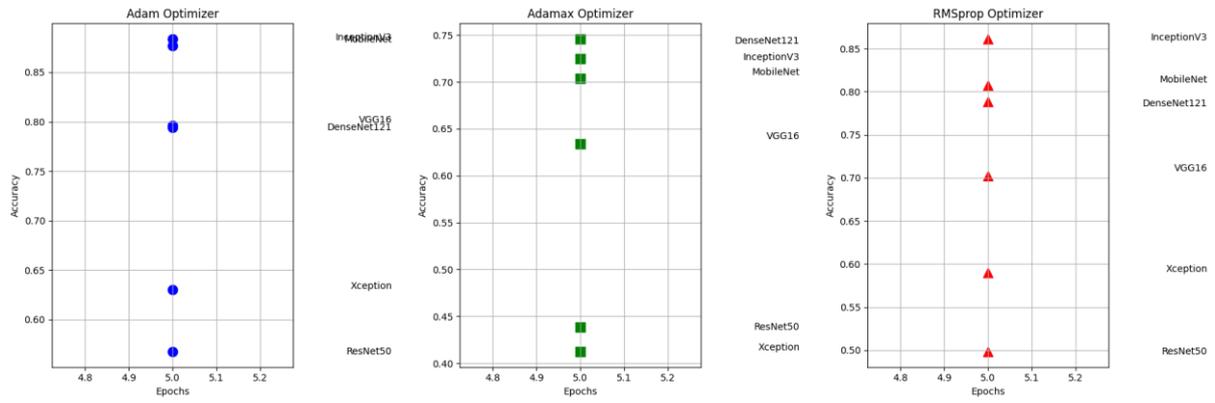

Figure 11: Transfer learning CNN models, accuracy, and epochs

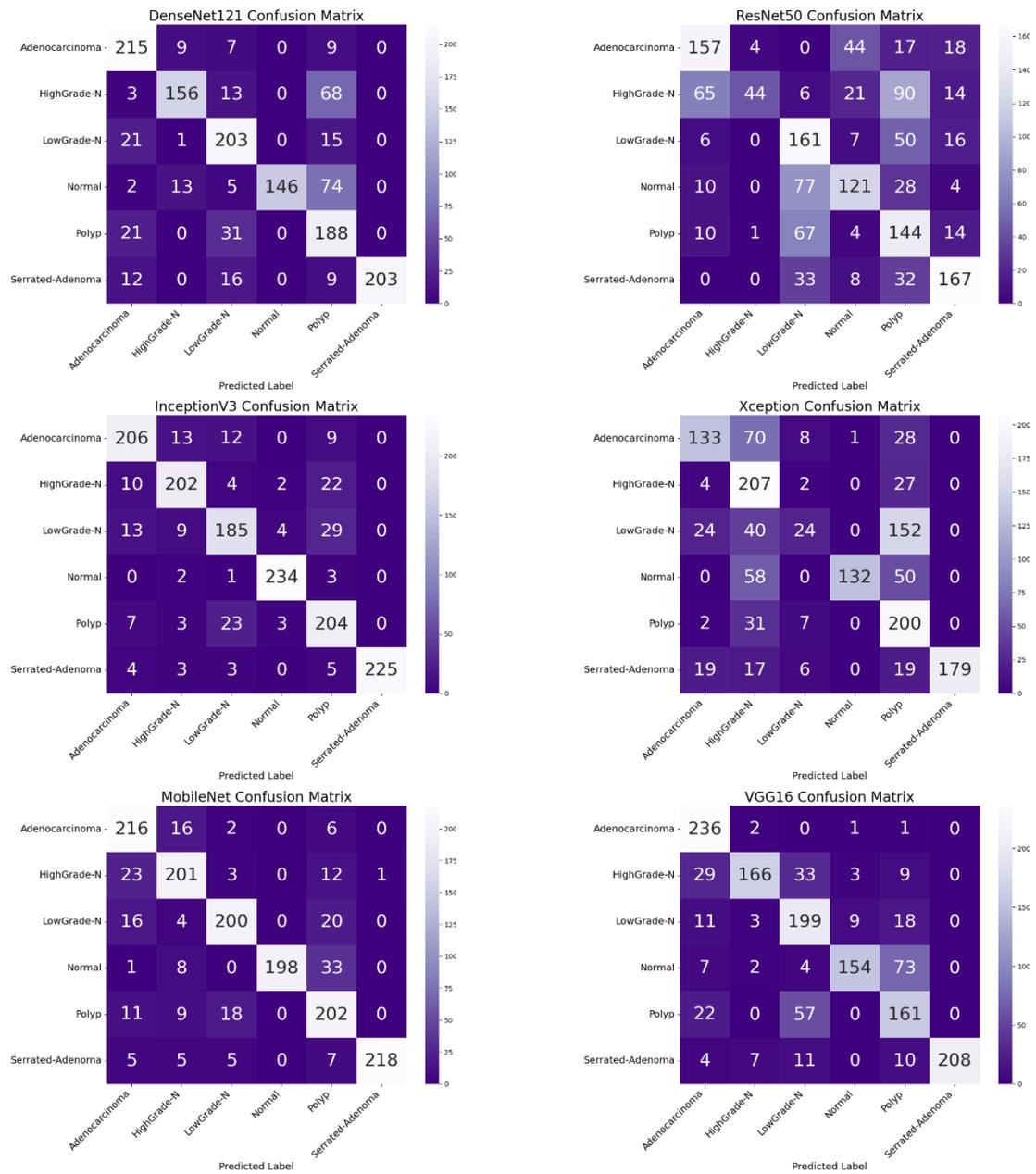

Figure 12: Confusion matrices of Transfer learning.

The classification results in Figure 12 show that Type 1 (false positives) and Type 2 (false negatives) errors are minimal for both DenseNet121 and MobileNetV2, thus demonstrating superior classification accuracy. The performance of Xception alongside InceptionV3 shows moderate errors of false positives and negatives. VGG16 exhibits failures in classification accuracy due to its higher error rates, particularly in distinguishing between the Normal and Polyp categories, thus demonstrating inferior reliability compared to competing models. ResNet50 demonstrates the highest degree of misclassification because it generates numerous false positive and false negative results that show its operational effectiveness.

## 4.4 Experiment 3: Ensemble Model Performance

Two ensemble models were built in this research project, with one model using DenseNet-Inception-Xception for Multi-path Depth-Width based and the other implementing ResNet50-InceptionV3-VGG16 for Multi-path Depth-Spatial based.

### 4.4.1 Ensemble 1

The comparison of multi-path-depth-width CNN ensemble (DenseNet–Inception-Xception) through Soft Voting, Hard Voting, and Rank-Based methods is presented in Table 6.

Table 6: Performance of Multi-path-depth-Width based (DIX) Ensemble.

| Multi path-depth-Width based | Ensemble method | Accuracy | Precision | Recall | F1 Score |
|---|---|---|---|---|---|
| DenseNet – Inception-Xception | Soft Voting | 98.02% | 98.12% | 98.02% | 98.07% |
| | Rank-Based | 57.19% | 65.71% | 57.19% | 59.43% |
| | Hard Voting | 95.52% | 96.13% | 95.52% | 95.53% |

Soft Voting Ensemble reaches 98.02% accuracy through minimal errors (Type I = 16, Type II = 16), which appear in the Soft Voting confusion matrix. The detection method demonstrates 98.12% precision, along with 98.02% recall, and achieves an F1 Score of 98.07%, which positions it as the most effective tool for CRC diagnosis, according to Table 6. The Hard Voting Ensemble demonstrates an accuracy of 95.52% and both Type I and Type II errors amount to 43 each. The Hard Voting confusion matrix shows 96.13% precision alongside 95.52% recall. The Rank-Based Ensemble generates only 57.19% accuracy alongside a high number of errors (Type I = 182 and Type II = 182) through the Rank-Based confusion matrix that creates poor performance with 65.71% precision and 57.19% recall using Table 6.

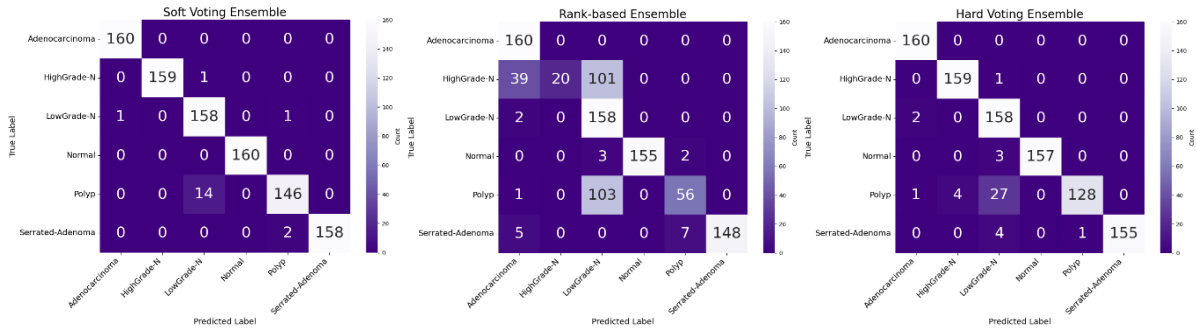

Figure 12: Confusion matrices of Transfer learning.

### 4.4.2 Ensemble 2

The comparison of the Multi-Path-Depth-Spatial based ensemble (ResNet50-InceptionV3-VGG16) through Soft Voting, Hard Voting, and Rank-Based methods is presented in Table 7.

Table 7: Performance of Multi-path-depth-Spatial based (RIV) Ensemble.

| Multi-Path-Depth-Spatial based | Ensemble method | Accuracy | Precision | Recall | F1 Score |
| --- | --- | --- | --- | --- | --- |
| ResNet50-InceptionV3-VGG16 | Soft Voting | 98.23% | 98.25% | 98.23% | 98.23% |
|  | Rank-Based | 89.69% | 89.83% | 89.69% | 89.71% |
|  | Hard Voting | 88.85% | 89.15% | 88.85% | 88.79% |

The Soft Voting Ensemble reaches a 98.23% accuracy rate and shows 11 Type I errors as well as 16 Type II errors in its Soft Voting confusion matrix. According to Table 7, the Soft Voting Ensemble represents the best method for CRC detection since it reaches 98.25% precision, 98.23% recall, and an F1 Score of 98.23%. The Hard Voting Ensemble reaches 88.85% accuracy, but it produces more errors than Soft Voting (Type I = 91, Type II = 107), as the Hard Voting confusion matrix reveals, with 89.15% precision and 88.85% recall. The Rank-Based Ensemble method demonstrates an 89.69% accuracy rate while producing errors consisting of 72 Type I and 84 Type II classifications according to the analysis of the Rank-Based confusion matrix (Table 7).

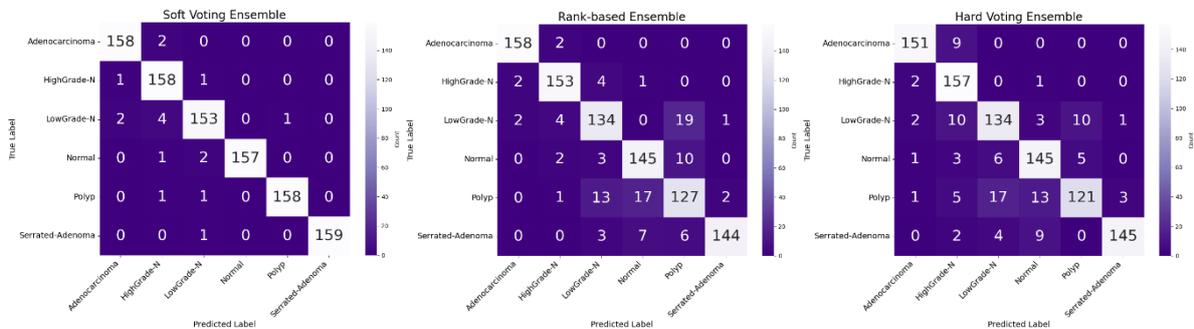

Figure 13: Confusion matrices of Transfer learning.

## 4.5 Result of the XAI

To provide both local and global explanations for the proposed R-Net model, this study employed three XAI methods: LIME, SHAP, and Grad-CAM.

### 4.5.1 LIME Visualization

As such, LIME is used for generating an explanation for each prediction. Figure 14 below shows that each explanation of LIME includes two features: Green and Red. "Green regions" are areas that positively contributed to the predicted class, and "Red regions" are areas that negatively contributed to the predicted class.

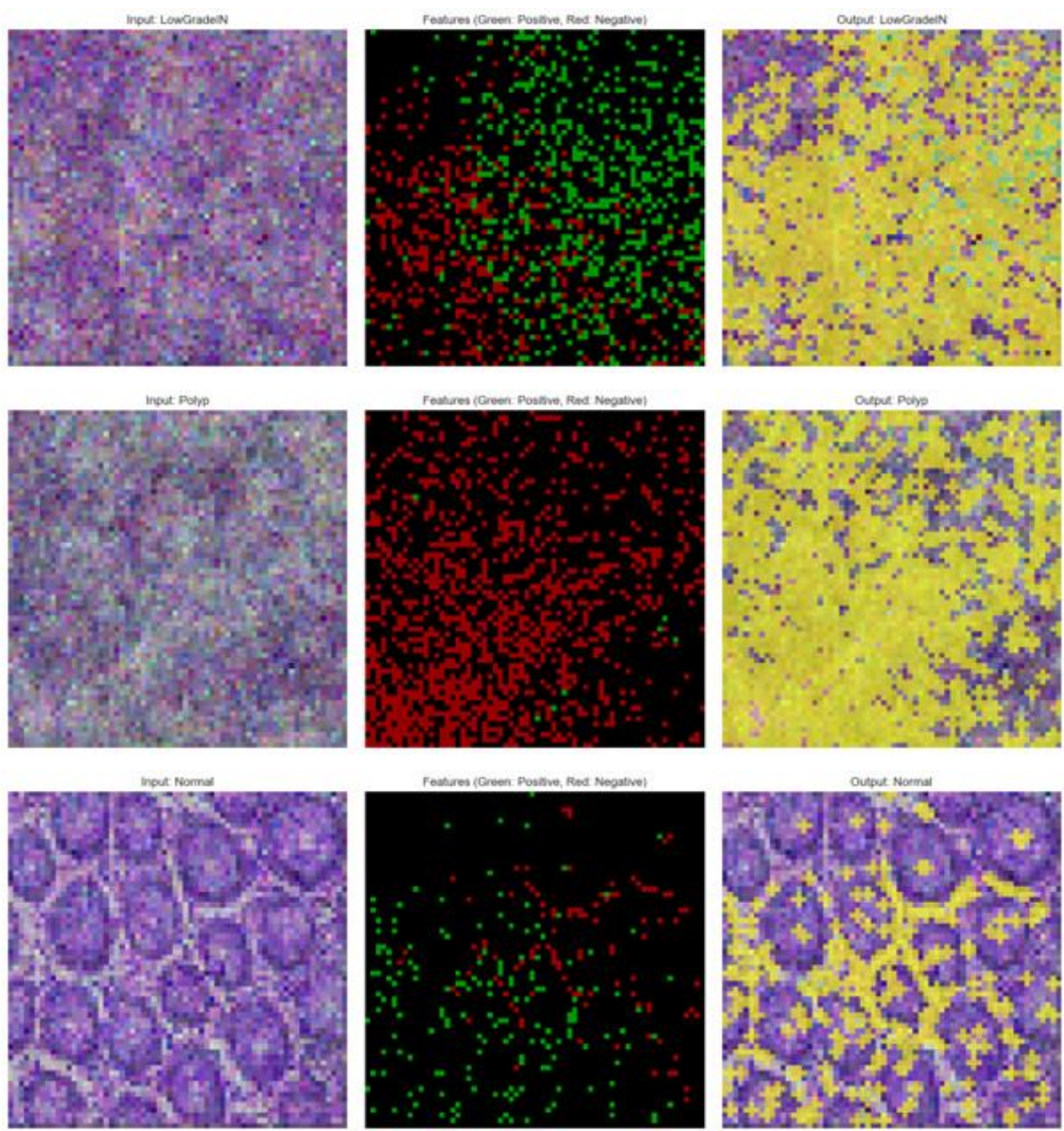

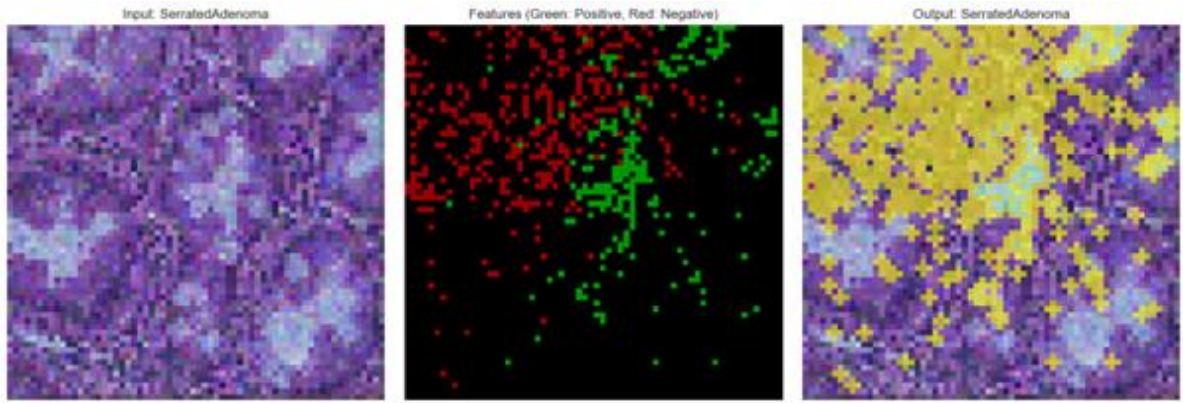

Figure 14: LIME explainer of CRC images

### 4.5.2 SHAP Explanation

Figure 15 shows each explanation of SHAP, including two features: Red and Blue. "Red regions" are areas that positively contributed to the predicted class, and "Blue regions" are areas that negatively contributed to the predicted class, along with a mean SHAP value for each prediction.

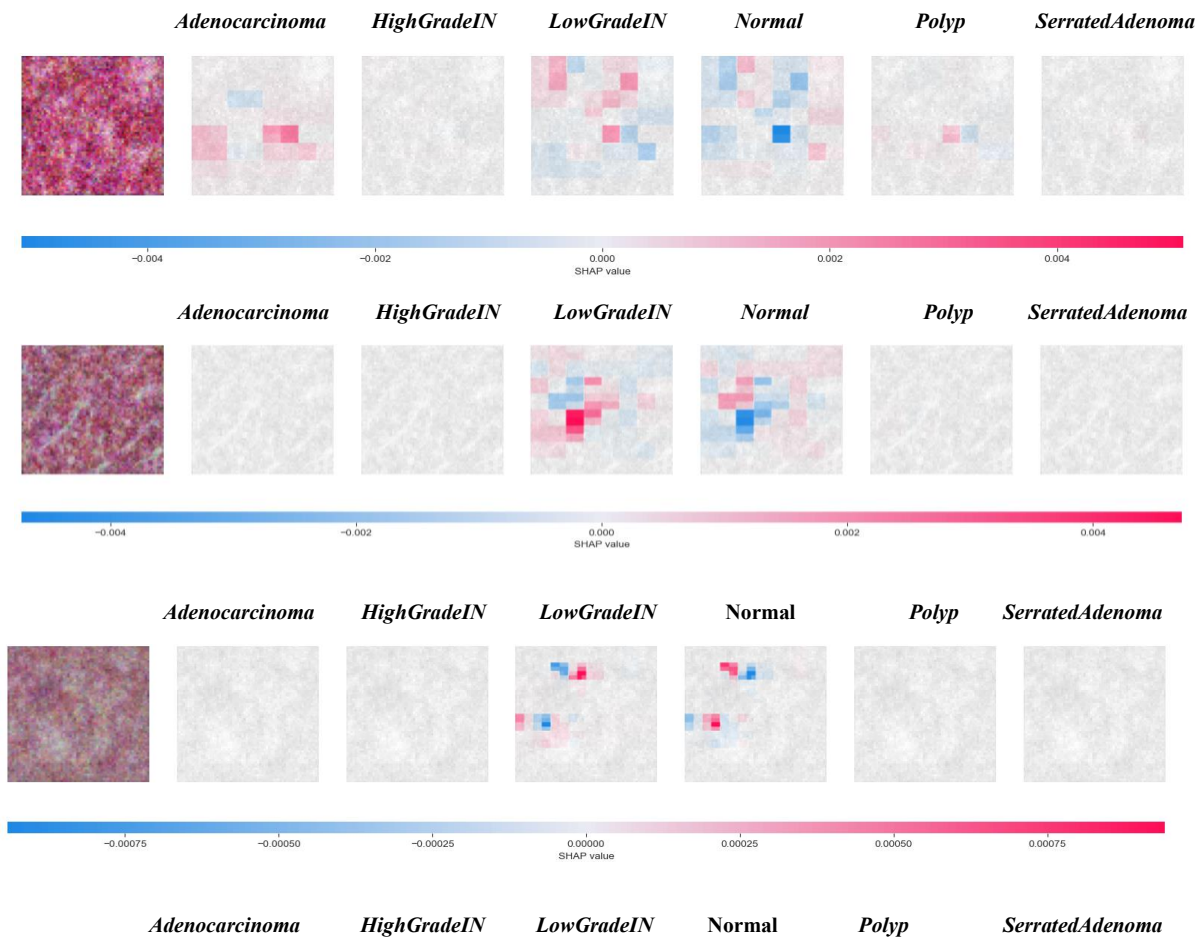

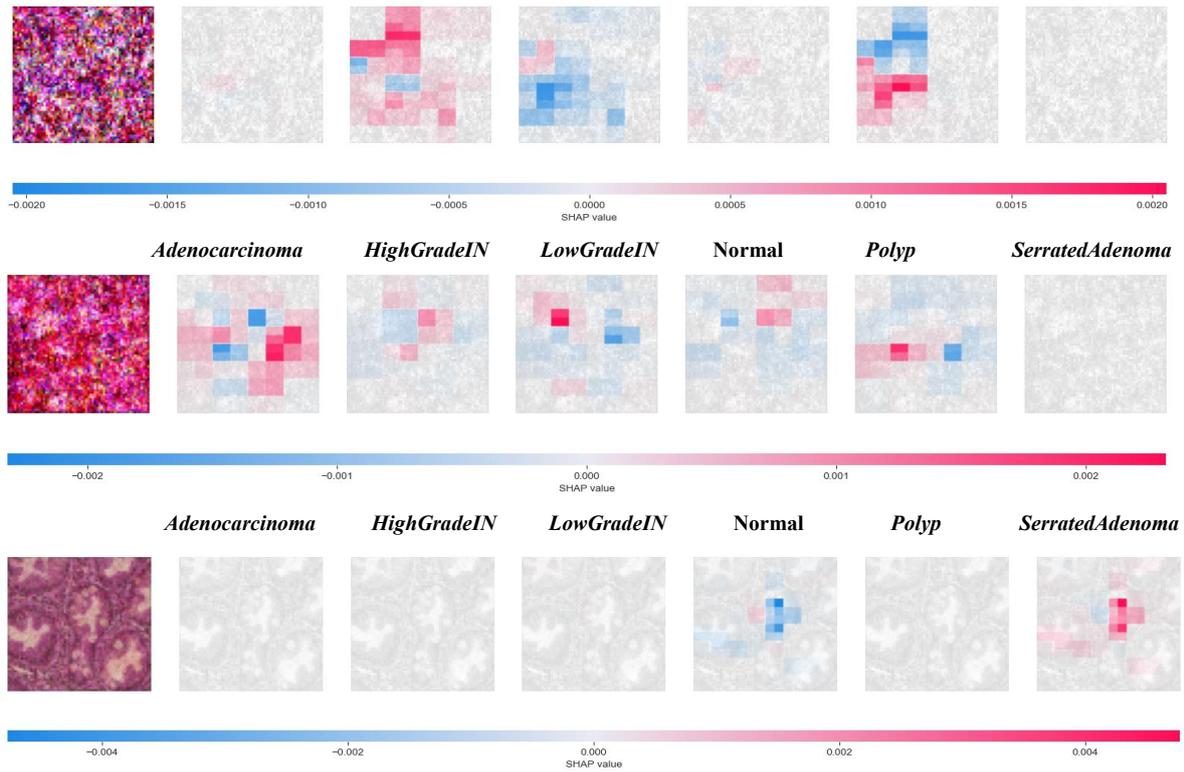

Figure 15: SHAP explainer of CRC images

### 4.5.3 Grad-CAM Analysis of Correctly/Incorrectly CRC Modalities

The Grad-CAM tool reveals which parts of an image the R-Net classifies most for predictions. Grad-CAM measures the connection between feature maps and class prediction by calculating the gradients within the last CNN layer. The heatmap shows how important each feature map is to the prediction results. The model employs gradient results to readjust its feature maps before combining them into the visualization map. The research employs Grad-CAM analysis to demonstrate which areas of the CRC image the model selected for diagnostic purposes. For example, Figure 16 shows the misclassification of images.

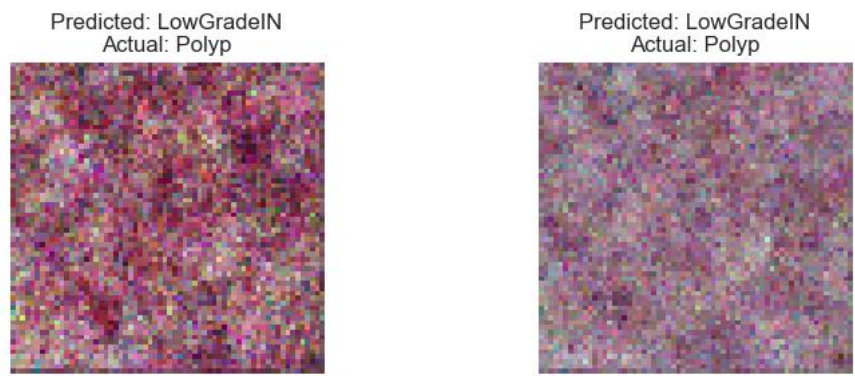

Figure 16: Examples of misclassified images

Grad-CAM technology displays which CRC cases the model incorrectly identified by showing where it focused its attention in Figure 17. The heatmaps show locations that received the highest attention through bright red and lesser attention with blue coloring. The visualizations demonstrate that the model examined areas of the colorectal that did not contain cancer. However, it made incorrect diagnoses, such as classifying cancerous cells (Polyps) without cancer as LowGradeIn.

Original image: Polyp; Predicted: LowGradeIn       Original image: Polyp; Predicted: LowGradeIn

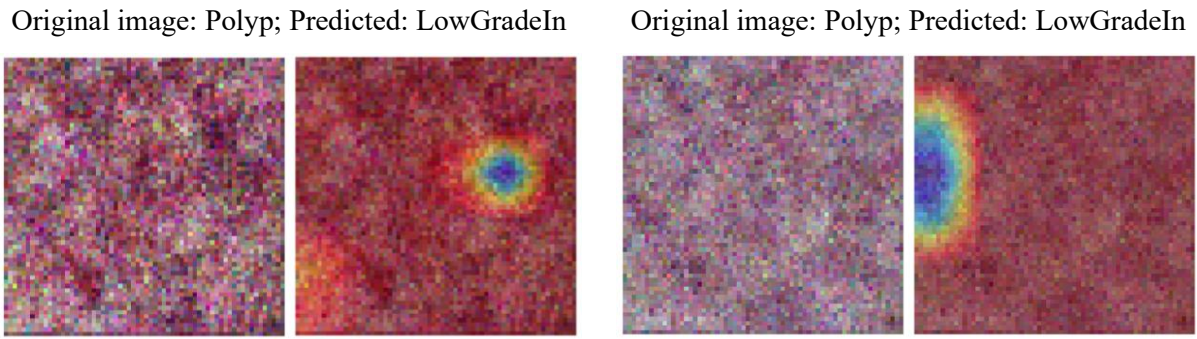

Figure 17: GRAD-CAM view of misclassified images

Figure 18 presents Grad-CAM highlighting the medical model's accurate recognition of tumor areas on correctly classified images. While we observe this behavior in specific non-tumor cases, our model tends to direct irrelevant attention to parts of the image, suggesting future development is needed in feature identification processes.

Original image: Polyp; Predicted: LowGradeIn       Original image: Polyp; Predicted: LowGradeIn

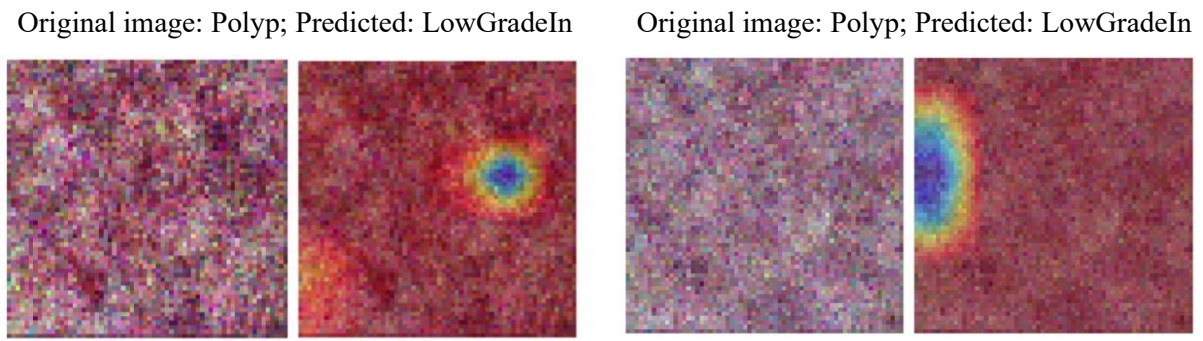

Figure 18: Grad-CAM view of correctly classified images

4.5.4 Pixel Intensity of Correctly/Incorrectly CRC Modalities

The pixel intensity displays feature attribution by showing which parts of the input images helped to make the CNN's decisions. This interactive graph displays aspects of a neural network model that misidentified a polyp as a low-grade cancerous cell (Figure 19). While Figure 20 shows the actual prediction of CRC. The panel displays the CRC images of the true cancer regions. The right panel demonstrates the Gradient × Input method that shows the model determination through pixel intensity values based on part contributions to the image. Parts of the image with intense colors had a significant impact on the prediction. The model selected non-cancerous areas for importance during classification because its Gradient × Input analysis did not match the cancerous regions. The mismatch between the model's learned features and the key characteristics of cancerous cells indicates that the model cannot provide an accurate assessment.

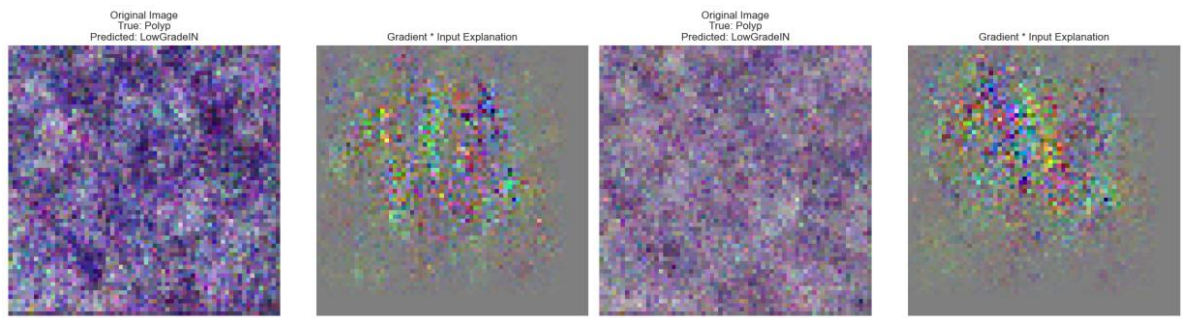

Figure 19: Grad-CAM view of pixel intensity for misclassified images

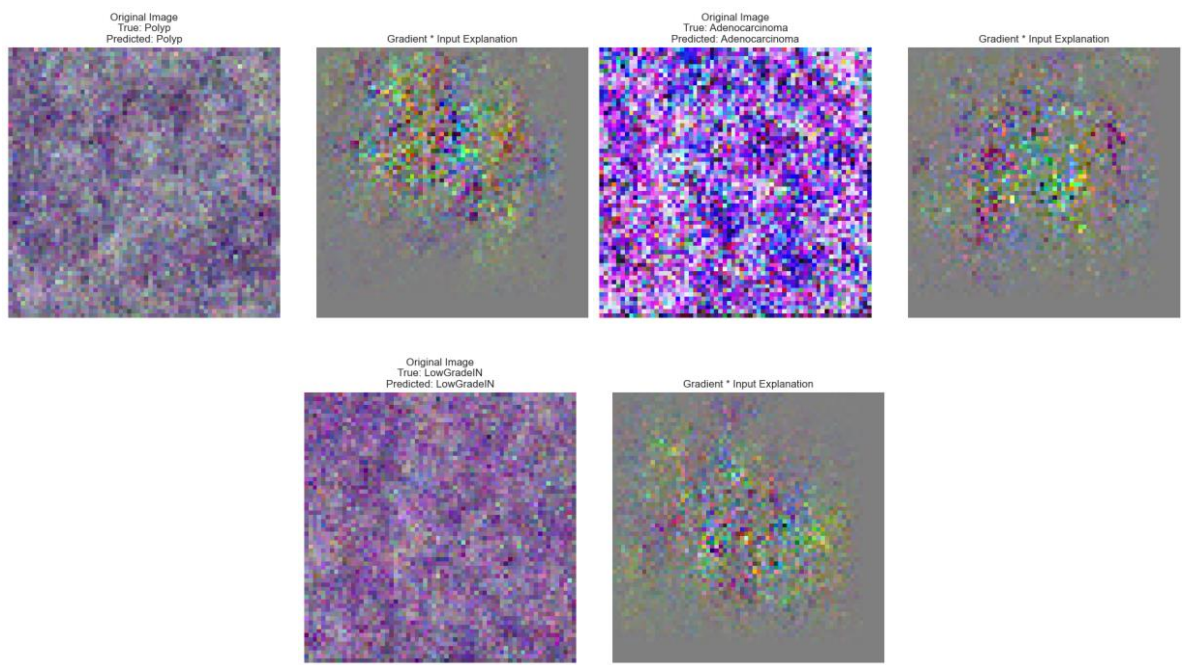

Figure 20: Grad-CAM view of pixel intensity for correctly classified images

## 5. Discussion

This research proposes R-Net (Reliable Net) as a compact CNN that effectively detects CRC through fewer layers while utilizing small learnable parameters. The proposed R-Net achieves a 99.37% accuracy in CRC image classification compared to SOTA CNNs, transfer learning, and ensemble models. Notably, the stable performance of Adam remains consistent, while RMSprop shows variable results between models, which proves that optimizer selection should consider the specific convergence patterns of the designed framework.

A state-of-the-art evaluation utilized six CNN architectural models, including InceptionV3, VGG16, MobileNet, ResNet50, DenseNet121, and Xception, for CRC classification. Both Xception and DenseNet121 yield equivalent diagnostic outcomes, but they require significantly more computational power than R-Net. Through transfer learning methods, InceptionV3 and MobileNet executed better classification than alternative approaches, whereas they did not match R-Net's efficiency levels. The combined utilization of multiple CNNs through ensemble models achieved high classification results, and Soft Voting proved to be the most effective method. However, R-Net proves to be a practical choice over ensemble methods because it delivers effective results while requiring less computational power and shorter training duration.

R-Net prediction validation and interpretability were improved by the implementation of XAI techniques, which included LIME SHAP together with Grad-CAM. Visualizations from Grad-CAM demonstrated that R-Net accurately detects cancerous regions, contributing to the accuracy of its diagnostic decisions. The detailed explanations provided by LIME and SHAP helped identify problematic predictions, thereby enhancing the trustworthiness of the model.

The performance evaluation of R-Net classification is presented in Figure 21, which compares the accuracy between individual CNNs, transfer learning models, and ensemble methods.

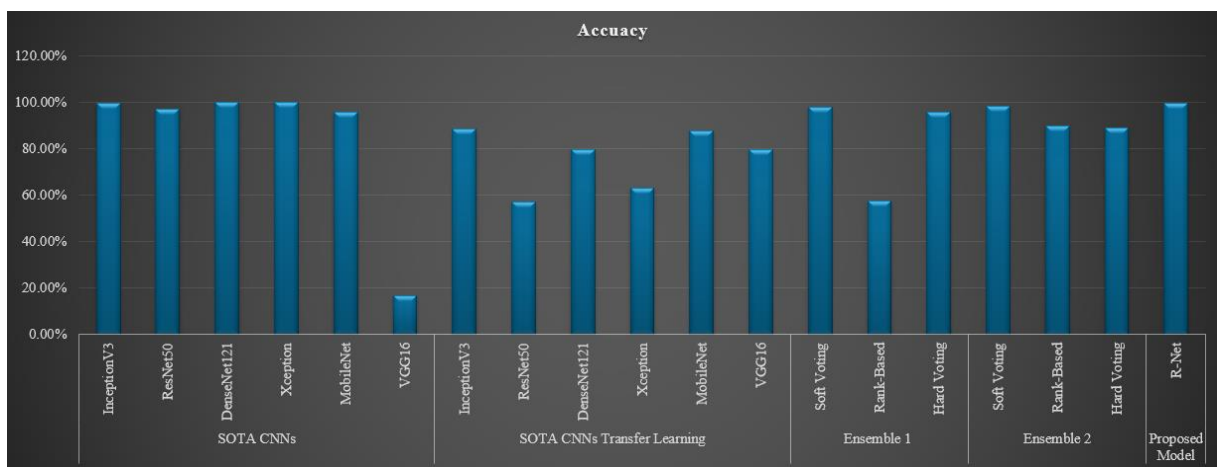

Figure 21: Accuracy comparison among individual CNN, transfer learning, and ensemble models.

The comparative performance results in Table 13 show that R-Net produces better outcomes than XGBoost, Ensemble, Transfer Learning, and Interpretable Machine Learning Systems by achieving 99.37% accuracy.

Table 13: Performance comparison of the proposed R-Net model with other models.

| Authors | Model | No. of Classes | Accuracy (%) |
|---|---|---|---|
| Georgiou et al., (2024) | XGBoost, Ensemble | 3 | 89.79% |
| Sirinukunwattana et al., (2022) | Consensus Molecular Subtype Classification | 4 | 92% |
| Neto et al. (2024). | Interpretable ML System | 4 | 94.5% |
| Kassani et al., (2022) | Transfer Learning | 4 | 95% |
| Yamashita et al. (2023). | DL for Microsatellite Instability Prediction | 3 | 97.3% |
| Elshamy et al., (2024) | Modified DNN Optimizer | 3 | 98% |
| Proposed Model | R-Net | 6 | 99.37% |

The research establishes R-Net as a highly accurate and efficient model which can perform CRC classification. R-Net proves suitable for medical use because of its robust combination of user-friendly interpretability with high performance capabilities, along with low system requirements. Future researchers will continue to develop the model further, as well as enhance data augmentation methods, while conducting rigorous clinical assessments to improve reliability in medical diagnostic contexts.

## 6. Conclusion and Future Work

The research investigates the effectiveness of DL models in CRC detection and classification by conducting three primary experiments. The researchers applied six CNN models, VGG16, ResNet50, DenseNet121, Xception, InceptionV3, and MobileNet, to analyze histopathological CRC images—secondly, the research utilized transfer learning techniques to enhance model classification results. The proposed R-Net achieved superior accuracy and efficiency while utilizing XAI techniques, including LIME, SHAP, and Grad-CAM. The R-Net showed enhanced reliability as its XAI framework delivered valuable insights about model prediction features and pixel intensity testing between correct and incorrect classifications.

The research study offers valuable results, but it also has some limitations. Using secondary datasets reduces the application range, which reveals the necessity of analyzing extensive, varied datasets for analysis. A wider test of the model was necessary because its training occurred exclusively through histopathological image analysis. Current research requires further investigation to establish the impact of transfer learning on lightweight CNN models, as the demonstrated results have not been promising. Medical expert confirmation serves as a requirement for models to acquire a credibility status. The evaluation of various program models is necessary to optimize their efficiency, performance, and adaptation capabilities before adoption for practical use.

In conclusion, study results demonstrate that CNN technology proves highly efficient in identifying CRC during screening examinations. The R-Net system achieves high accuracy in medical image classification through its practical and lightweight structure, which protects readability. Modern research must link healthcare professionals with advanced imaging technology usage to enhance both DL CRC diagnosis detection methods and clinical diagnostic capabilities.